\documentclass[letterpaper]{article} 
\usepackage{aaai2026}  
\usepackage{times}  
\usepackage{helvet}  
\usepackage{courier}  
\usepackage[hyphens]{url}  
\usepackage{graphicx} 
\usepackage{natbib}  
\usepackage{caption} 
\usepackage{algorithm}
\usepackage{algorithmic}
\usepackage{booktabs}
\usepackage{multirow}
\usepackage{fncylab}
\usepackage{multicol}
\usepackage{listings}

\makeatletter
\newcommand{\setlabel}[1]{\edef\@currentlabel{#1}\label}
\makeatother

\usepackage{newfloat}
\usepackage{listings}
\DeclareCaptionStyle{ruled}{labelfont=normalfont,labelsep=colon,strut=off} 
\floatstyle{ruled}
\newfloat{listing}{tb}{lst}{}
\floatname{listing}{Listing}
\title{A Framework for Testing and Adapting REST APIs as LLM Tools}

\author{
    Jayachandu Bandlamudi, Ritwik Chaudhuri, Neelamadhav Gantayat, Sambit Ghosh, \\
    Kushal Mukherjee, Prerna Agarwal, Renuka Sindhgatta, Sameep Mehta
}
\affiliations{
    IBM Research, India\\

}

\usepackage{bibentry}

\begin{document}

\maketitle

\begin{abstract}
Large Language Models (LLMs) are increasingly used to build autonomous agents that perform complex tasks with external tools, often exposed through APIs in enterprise systems. Direct use of these APIs is difficult due to the complex input schema and verbose responses. Current benchmarks overlook these challenges, leaving a gap in assessing API readiness for agent-driven automation. We present a testing framework that systematically evaluates enterprise APIs when wrapped as Python tools for LLM-based agents. The framework generates data-aware test cases, translates them into natural language instructions, and evaluates whether agents can correctly invoke the tool, handle their inputs, and process its responses. We apply the framework to generate over 2400 test cases across different domains and develop a taxonomy of common errors, including input misinterpretation, output failures, and schema mismatches. We further classify errors to support debugging and tool refinement. Our framework provides a systematic approach to enabling enterprise APIs as reliable tools for agent-based applications.
\end{abstract}

\section{Introduction}
Recent advances in Large Language Models (LLMs) have enabled agents that can reason, plan, and call external tools to autonomously execute tasks~\cite{yao2023react, xu2023rewoo}. Tools allow agents to interact with external systems and data sources~\cite{masterman2024landscape}, making them central to enterprise automation. In practice, these tools are often implemented as REST APIs, which expose the functionality of enterprise software. However, REST APIs are typically not designed with LLM agents in mind, creating challenges for seamless integration due to complex input schema and verbose responses.  

Existing benchmarks, such as the Berkeley Function Calling Leaderboard (BFCL)~\cite{berkeley-function-calling-leaderboard}, $\tau$-Bench~\cite{taubench}, and ToolACE~\cite{toolace2024}, highlight challenges in LLM tool use, including incorrect API selection, malformed arguments, and hallucinated usage. However, benchmarks are limited in tool diversity, domain coverage, and complexity, restricting their utility for evaluating enterprise readiness. In contrast, our approach focuses on strengthening the tools themselves by reusing mature testing methods to assess tool robustness.  

Traditional testing frameworks such as Postman\footnote{\url{https://www.postman.com/}}, RESTler~\cite{atlidakis2019restler}, and EvoMaster~\cite{arcuri2021evomaster} generate and execute tests from API specifications, validating correctness and robustness under diverse conditions~\cite{wu2019automated}. Recent work also explores LLMs for automated test generation~\cite{llmtestcase2024}. While effective for conventional validation, these approaches do not evaluate whether APIs are \emph{agent-ready}—that is, usable as tools by LLM agents.  

Our work bridges this gap by adapting established testing techniques to assess REST APIs as tools for LLM-based agents. We transform REST APIs as agent framework specific Python tools, and further transform conventional test cases into natural language utterances that mimic the inputs agents receive, enabling evaluation of the agent’s ability to interpret instructions, invoke the correct tools, and process outputs. 


Specifically, our contribution is the creation of a novel framework to automatically evaluate the compatibility of REST APIs as tools for LLM-based agents, which includes:  
\begin{itemize}
    \item We generate diverse test cases using established methods, ensuring coverage of different input scenarios and expected behaviors.  
    \item We develop an LLM-based approach to transform test cases into natural language utterances, and evaluate whether agents can map them to valid tool calls and interpret tool responses.  
    \item We construct a taxonomy of common error patterns from 2400 tool test case executions across six enterprise applications, providing actionable insights for improving tool definitions.  
\end{itemize}  

\begin{figure*}[ht]
\centering
\includegraphics[width=0.8\textwidth,height=7cm]{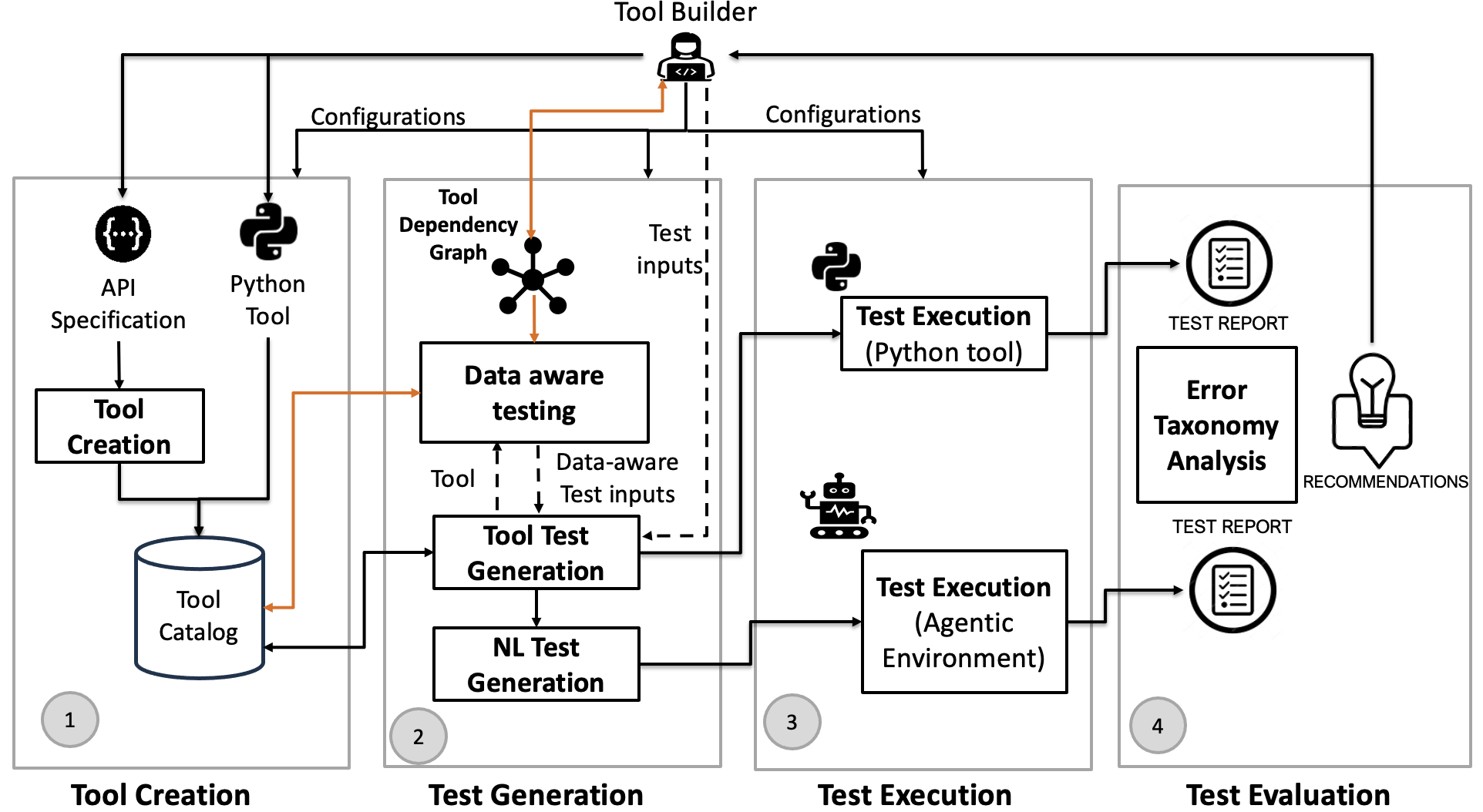}
      \caption{Framework for Tool Testing}
       \label{fig:framework}
\end{figure*}
\section{Related Work}

Stand-alone API testing is well-studied in the literature. API specifications~\cite{oapi2023} are largely machine-readable, but they also allow developers to include natural language descriptions in certain fields. A number of automated testing tools such as EvoMasterBB~\cite{arcuri2019restful}, bBOXRT~\cite{laranjeiro2021black}, RESTest~\cite{martin2021restest}, and RestTestGen~\cite{corradini2022automated} have been developed. However, these rely more on the machine-readable parts of API specifications and do not leverage the human-readable descriptions. Advances in large language models, recent literature~\cite{zhao2023survey} demonstrates that the human-readable part of the API specifications, which includes descriptions, titles, and examples, can be used to construct coherent test cases, keeping in mind the dependency between the parameters~\cite{martin2019catalogue}. 


While stand-alone API testing is well studied, testing of tools as a part of an agentic framework remains relatively unexplored. Various benchmarks are available for assessing tool utilization, based on the accuracy of tool selection (e.g., Gorilla~\cite{gorilla2023}) and based on the response generated (e.g., ToolQA~\cite{zhuang2023toolqadatasetllmquestion}). T-Eval~\cite{chen2024tevalevaluatingtoolutilization} offers a systematic evaluation framework by decomposing tool utilization into discrete sub-processes like reasoning and planning, providing fine-grained insights into the tool utilization capabilities of LLMs. EasyTool~\cite{yuan2024easytoolenhancingllmbasedagents} introduces a framework that streamlines and standardizes tool documentation, enabling LLMs to process diverse and redundant instructions by transforming them into concise and unified interfaces, which improves performance in real-world tasks.

This work builds on prior research in API testing and extends it to the evaluation of REST APIs when they are wrapped as Python tools and integrated into agentic systems.

\section{Automated Tool Testing Framework}

The framework for automating tool testing in an agentic flow is presented in Figure~\ref{fig:framework}. The process begins with the API specification provided by the Tool Builder. The specification is first transformed into a functional \textit{tool} compatible with the agentic framework and stored in a tools catalog, which serves as a repository of callable resources during testing. If the \textit{tool} is readily available, the Tool Builder can add it to the catalog.

Next, multiple test cases are automatically generated using an LLM-based test case generation approach to ensure comprehensive coverage. Additionally, a Tool Dependency Graph can be generated for the tool catalog to capture dependencies among tools and provide data-aware test inputs (e.g., a valid \texttt{user\_id} or a system-specific \texttt{location} code). The Tool Builder reviews the Tool Graph and decides whether to use it for obtaining data-aware test inputs. For each test case, several Natural Language (NL) test cases are generated, which are used as utterances to test tools in the agentic environment. Both the generated test cases and the tool serve as inputs to subsequent phases.

Test execution proceeds along two paths. Test cases are executed directly by invoking the corresponding Python tool with their specified inputs. In parallel, NL test cases are executed within the agentic environment. The framework instantiates ReAct-style agents~\cite{yao2023react}, which process the natural language input, select tools, and execute tool calls. In each phase, the Tool Builder can provide necessary configurations such as the number of test cases, test input data, choice of LLM model, and tool access credentials.

Finally, in the Tool Evaluation phase, test reports from direct Python tool execution and agent-based tool execution are compared. The direct tool execution report is treated as ground truth, while the NL test execution report reflects the agent's behavior. Discrepancies are analyzed to identify common error types based on a taxonomy. This analysis provides actionable recommendations for tool robustness.

\subsection{Tool Creation}\label{sec:tool-creation}
The process begins with an API specification, typically provided in the OpenAPI Specification (OAS) format~\cite{oapi2023}, which defines endpoints, HTTP methods, parameters, and responses. Each path operation in the OAS is converted into a functional tool. We parse the specification to obtain the required information for creating a `tool definition', which is a Python-executable LangChain\footnote{\url{https://python.langchain.com/docs/concepts/tools/}} tool.\\
Each tool definition is instantiated with a set of structured attributes derived from the OAS. The tool \textit{name} is taken directly from the endpoint `operationId' (e.g., getApiV2Tickets). In the tool definition, a method is created with the tool name to call the  endpoint URL. For method inputs, we parse the API specification to extract path parameters, query parameters, and request-body fields, together with their associated metadata such as data types, required/optional flags, and constraints. Similarly, we parse the endpoint responses to construct the method output schema. The tool \textit{docstring} is constructed from endpoint and parameter descriptions, explicitly marking required and optional fields. 
To guide the LLMs towards correct parameter usage, the docstring is enhanced with example values available in the OAS. Tool authentication is created using the OAS \textit{headers} information, with authentication credentials populated from the Tool Builder configuration. An example of the tool is presented in the Appendix (Figure~\ref{fig:kubernetes-tool}).

\subsection{Test Case Generation}\label{sec:test-case-generation}

We employ LLMs for test case generation~\cite{llmtestcase2024} and extend the work by integrating additional constraints extracted from the tool definition. The \textit{tool definition} details the tool’s signature along with its input and output schema definitions, and the docstring containing tool and parameter descriptions. The LLM is prompted for automated generation of diverse test cases that achieve broader coverage. To this end, we define three types of scenarios: i) test cases with only mandatory parameters, ii) test cases with all parameters, and iii) test cases with mandatory and subsets of optional parameters adhering to interdependency constraints of parameters. The subset of the optional parameters are chosen by LLM during the test case generation. Leveraging an LLM for scenario selection allows for more complex and varied parameter combinations, ensuring diverse coverage. The overall process proceeds in two stages:

\paragraph{Step 1: Parameter information and constraint extraction.}
We first extract parameters and their associated constraints using a prompting strategy similar to that in~\cite{llmtestcase2024}. The model is instructed to identify parameters by name and description and to extract corresponding constraints. These include: i) \emph{type and format constraints} (e.g., type: array, format: date), ii) \emph{value constraints} which restrict permissible ranges (e.g., maximum values), iii) \emph{example values} (e.g., country or currency codes), and iv) \emph{operation constraints}, which define dependencies among parameters. These extracted constraints enrich the original tool specification to guide subsequent test generation.

\paragraph{Step 2: Constraint-aware parameter value generation.}
We then provide the LLM with a prompt containing the tool specification, parameter information, extracted constraints, nested parameter formats, and the type of test case scenario to generate. The model is instructed to output test cases in a structured key–value format, where each key corresponds to a parameter name and each value is an assigned parameter value. Depending on the scenario (mandatory-only, all parameters, or subsets of parameters), the LLM generates one or more valid test cases that conform to the specified constraints. The number of test cases generated is user configurable, with a default of 20. The generated test cases collectively cover all scenarios, including variations with all parameters, mandatory-only parameters, and subsets of parameters. Sample prompts for test case generation are provided in Appendix(\ref{appendix:testcasegenprompt}). 

\paragraph{Incorporating custom test inputs.}
When predefined test data inputs are available for a given tool specification, we incorporate them into the process. These predefined data values for parameter instances are left unchanged unless they lack values for mandatory parameters. In such cases, we prompt the LLM to fill in the missing values by combining the incomplete test data with the tool. Predefined test data inputs are particularly valuable for producing relevant test cases in data-aware testing.

Through this process, we generate structured test cases for each tool, which are then transformed into natural language test cases for execution in automated tool testing.

\subsection{Data Aware Test Input Generation using Tool Dependency Graph}\label{sec:data-aware-test-case-generation}

LLM-based test case generation offers scalability and diversity but often operates as a black-box process. Many API-wrapped tools require input values that are available only at run time (such as valid \texttt{user\_id} or \texttt{account\_num}), which cannot be detailed in the API specifications. As a result, purely LLM-generated test cases may be syntactically valid but contain irrelevant data values, leading to empty responses, trivial outputs, or `no results found' errors (e.g., 404 API response code). In some instances, due to inter-tool dependencies, tools may require input values that originate from the outputs of other tools. LLM-generated test cases struggle to capture these dependencies, further limiting their effectiveness in real-world testing scenarios. 

To address these shortcomings, we introduce \emph{data-aware testing} that augments LLM-based generation with knowledge of enterprise-specific data obtained by leveraging a \textit{tool dependency graph}. By making the LLM aware of data formats and the semantic role of each parameter, the framework produces test cases that are syntactically correct, contextually grounded in relevant enterprise data, resulting in meaningful outputs from the target tools. The proposed approach is as follows:

\paragraph{Step 1: Tool dependency graph.}
To generate test cases with relevant input data for a given target tool, we first determine the ordered sequence of tools that culminates in the target tool. This sequence is obtained by identifying parent tools whose output parameters are consumed as inputs by the target tool, thereby forming a dependency graph that captures the full lineage of data. An LLM is prompted (Appendix:~\ref{ref:tool_graph}) to assist in inferring such dependencies. In the prompt, we provide tool docstrings as context to identify whether a dependency exists between a pair of tools. Similarly, we identify dependencies among all tools in the catalog to create a `tool dependency graph' with tools as nodes and the identified dependencies as edges.

\paragraph{Step 2: Parameter mapping in tool dependency graph}
Once the tool dependency graph is established, we align the \emph{output parameters} of parent tools with the \emph{input arguments} of the child tool. Since tools often vary in naming conventions, data formats, and even in the semantic interpretation of parameters, a direct mapping is rarely straightforward. To address this, we leverage an LLM to identify parameter mappings between pairs of tools (Appendix:~\ref{ref:argument_mapping}). However, because the dependencies and mappings depend heavily on the parameter names and descriptions provided by the tool builder, there will be inconsistencies. In some cases, tool definitions do not contain detailed docstrings, which can cause failures in identifying tool dependencies. Therefore, the generated tool dependency graph with parameter mappings is verified by the Tool Builder. After approval from the Builder, the tool dependency graph with parameter mappings is persisted in the tool catalog. 

\paragraph{Step 3: Tool sequence execution.} 
With a verified tool dependency graph and parameter mappings, we obtain the required tool sequence for the target tool by identifying all dependencies. Now we execute the tools sequentially along the dependency chain. At each stage, mapped outputs from parent tools are transformed into valid inputs for the next tool. When the execution reaches the step immediately preceding the target tool, the transformed outputs from the parent tools are captured. These outputs, that satisfy the target tool specifications, are formalized as \emph{data-aware test inputs} and utilized in tool test case generation.

\subsection{Natural Language Test Case Generation}\label{NL_testcase_generation}

The generation of natural language (NL) test cases involves creating coherent, human-like sentences to invoke and evaluate tools in an agentic environment. These utterances are crafted for each test case based on the provided tool definitions and associated input payloads.

We experimented with the following two approaches to generate the natural language test cases.

\textbf{Prompted LLM:} The first approach constructs a prompt by providing the tool’s function definition, including its name, description, input/output parameters, and test payload, as context. The prompt explains the tool's functionality, the required input parameters, and any constraints or conditions that are used to generate the NL utterance. The sample prompt is shown in the Appendix(\ref{appendix:nlgenprompt}).

We observed that the natural language (NL) test case utterances generated by the prompted model often lacked natural, human-like qualities and appeared robotic. For instance, the utterance would directly list the parameters in the test payload, which does not reflect the typical language or phrasing a user would employ in real-world scenarios. To alleviate this issue, we developed a second approach to create additional NL test cases. 

\textbf{Fine-tuned LLM:} The second approach is based on parameter-efficient fine-tuning (e.g., Low-Rank Adaptation LoRA ~\cite{lora2022}) of a small language base model~\cite{schick2020s}. The dataset for fine-tuning is derived from the Berkeley Functional Calling Leaderboard (BFCL) V3 dataset~\cite{berkeley-function-calling-leaderboard}.  The BFCL dataset is designed to benchmark large language models in function-calling tasks. Inputs consist of natural language queries and desired function specifications (similar to an API specification), while outputs include structured JSON objects representing the correct function name, parameters, and values. To fine-tune our model for the natural language generation task, we used the function specification along with structured JSON objects of parameters and values as input (in the same prompt format employed in Prompt LLM) and considered the natural language queries as the output.

\textbf{LLM as a judge:} To evaluate each NL test case, we employed a prompted LLM as a judge~\cite{zheng2023judging}, assessing them according to two criteria.

\noindent\textbf{Accuracy} (rating: Accurate, Partially Accurate, Inaccurate): Determine how well the utterance reflects the information in the sample payload, i) \textit{Accurate}: All values are correctly and completely reflected from the payload, with no missing or incorrect information. ii) \textit{Partially Accurate}: Some values are missing or slightly incorrect, but parts of the payload are accurately captured. iii) \textit{Inaccurate}: Major information is missing, incorrect, or unintended details are present.

\noindent \textbf{Naturalness} (range 1 to 5): Assess how natural and human-like the utterance sounds. Higher scores should be given to utterances that resemble natural spoken language and avoid synthetic, mechanical, or overly structured phrasing. Examples of natural and synthetic utterances are provided in the Appendix(\ref{appendix:NLExamples}).

The LLM assigns scores based on the provided in-context examples. A sample prompt for the LLM as a judge is given in Appendix(\ref{appendix:NL-LaJ}).

\subsection{Test Case Execution}
The generated test cases are executed to evaluate the functionality and performance of each tool under different scenarios. We first execute the test cases directly by invoking the corresponding Python tool with the specified input payload. The resulting outputs serve as the ground truth for validating tool behavior.

To assess test cases in an agentic environment, we follow a two-step process. In the first step, we create the environment by selecting tools from the same application and binding them within the agentic flow. Specifically, we choose a subset of $k(=10)$ tools whose descriptions are most similar to those of the target tool. This setup ensures that the agent can distinguish among related tools while reflecting a realistic usage context.

In the second step, each test case is invoked through its natural language (NL) utterance. The agent interprets the utterance, identifies the appropriate tool, and executes it accordingly. All execution events are recorded as NL test case reports. These reports are then compared against the ground truth outputs obtained from direct Python execution, verifying that the NL test cases produce the expected results. Importantly, the generated test cases are agent-agnostic and can be executed across different agent frameworks~\cite{yao2023react} and with various LLMs.

\subsection{Error Analysis and Action Recommendation}
When executing tools in an agentic environment, errors can arise for a variety of reasons. Existing function-calling benchmarks such as BFCL~\cite{berkeley-function-calling-leaderboard} and $\tau$-bench~\cite{taubench} categorize common issues, including function mismatches, missing parameters, and parameter type errors. Building on these efforts, our framework expands the taxonomy of error causes to include additional scenarios observed in practice, and leverages this taxonomy not only to characterize failures but also to recommend corrective actions.

\subsubsection{Error Taxonomy}
The goal is to create a classification system to help understand the types of errors that occur when an agent attempts to use external tools. The taxonomy is flexible enough to accommodate new types of errors as our experience with tool-calling expands. We use a qualitative approach based on analyzing tool execution events for the root causes of these errors. The taxonomy is presented in Figure~\ref{fig:taxonomy}. The errors identified are categorized into four main categories:
\begin{figure}[ht]
\centering
     \includegraphics[width=0.48\textwidth,height=5cm]{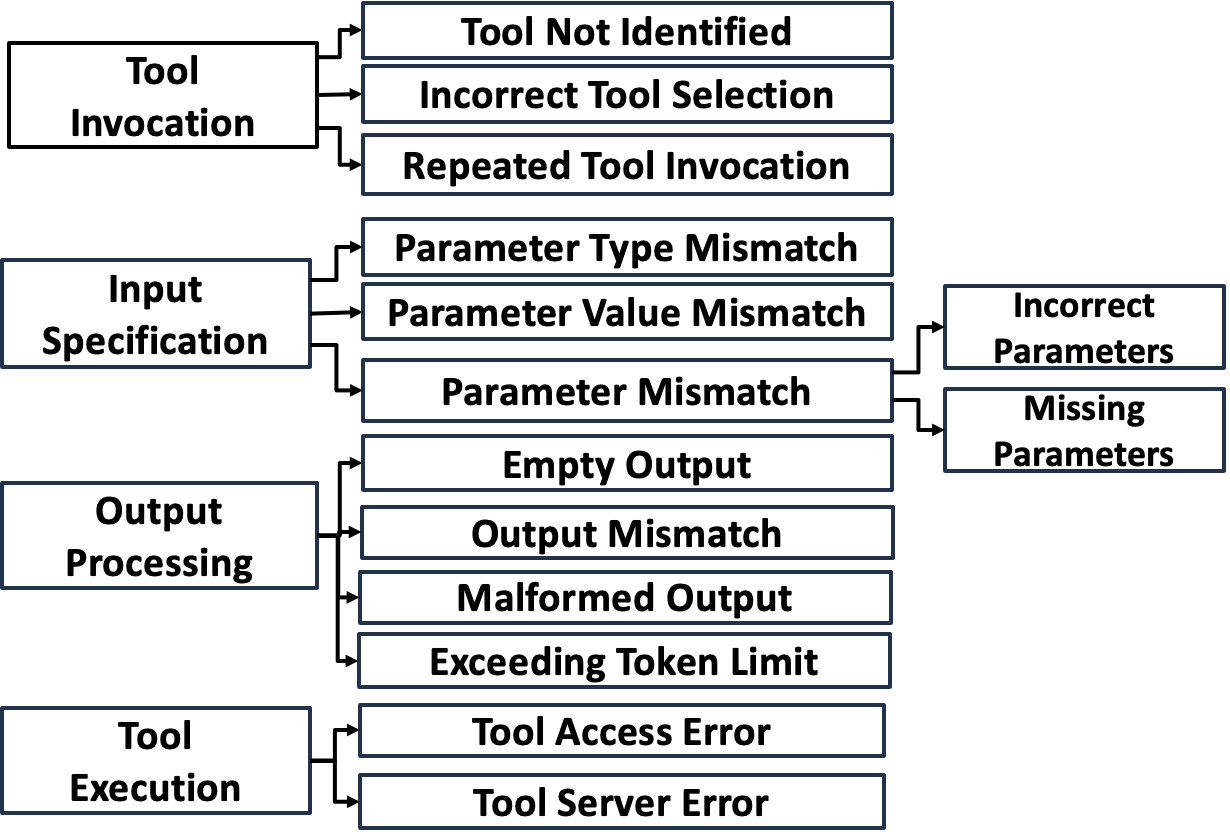}
      \caption{Error Taxonomy}
       \label{fig:taxonomy}
\end{figure}
\noindent \paragraph{\textbf{Tool Invocation Errors:}} These errors arise from the identification, initiation, or sequencing of tool calls, rather than from the tool’s internal functionality.
\begin{itemize}

\item\textbf{Tool Not Identified:} The agent fails to map the user’s intent to any available tool, leaving the task unexecuted.

\item\textbf{Incorrect Tool Selection:} The agent maps the user’s intent to the wrong tool, resulting in an inappropriate call. For example, it may invoke \texttt{putV2NotesById} instead of \texttt{getV2Notes} for an utterance such as “notify me for all notes tagged as household.”

\item\textbf{Repeated Tool Invocation:} The agent repeatedly calls the same tool without progress, often entering a loop when a tool returns uninformative responses (e.g., “Bad request,” “Missing body”). This behavior reflects the agent’s inability to determine a valid next step, resulting in resource exhaustion.
\end{itemize}

\paragraph{\textbf{Input Specification Errors:}}  These errors arise from issues in how the agent specifies the input parameters required by the tool.
\begin{itemize}
    \item \textbf{Parameter mismatch:} There can be two types of errors: i) \textbf{Missing parameters: }The agent omits one or more required parameters when calling the tool. The root cause is the agent's inability to determine all necessary parameters from its understanding of the task and the tool’s definition. ii) \textbf{Incorrect parameters:} The agent hallucinates parameter names by not considering the tool definition. For example, the agent creates a boolean parameter of name \texttt{show\_assigned\_departments} instead of \texttt{cansee\_assigned\_department}. Enterprise APIs may have non-intuitive parameter names
    \item \textbf{Parameter type mismatch:} The agent provides input parameters with data types that do not match what the tool expects. This might occur due to malformed extraction of input parameters or type mismatches from the underlying LLM. This is common when the input value is a single-item list. For example, passing `Image' as a string when [`Image'] needs to be passed.
    \item \textbf{Parameter value mismatch:} The agent may provide an incorrect parameter value despite having the utterances with correct values. There can also be scenarios where the LLM processes the input from the utterance. For example, `john doe' in the utterance is passed as `John Doe'.
\end{itemize}

\paragraph{\textbf{Output Processing Errors:}} These errors arise when the agent is parsing the output of the tool
\begin{itemize}
    \item \textbf{Output Mismatch:} Tool execution output may not contain all the necessary information according to the specified tool output schema. Any missing parameters from the expected output are identified as output mismatch. 
    \item \textbf{Empty Output:} Tool execution output sometimes will be empty with no results and conforms to the specified tool output schema. We observe this situation when there are no results found for the given inputs. 
    \item \textbf{Malformed Output:} The agent cannot parse or interpret the tool’s output, typically because the output is malformed or incorrect format for the agent to handle. 
    \item \textbf{Tool output exceeding LLM token limit:} The agent cannot process the output because the size of the response exceeds the token limit of the LLM.
\end{itemize}

\paragraph{\textbf{Tool Execution Errors:}} These errors arise when the tool is executing
\begin{itemize}
   \item \textbf{Tool access errors:} The agent cannot successfully execute the tool due to incorrect or insufficient access permissions with the response code of $4XX$(e.g., authentication failure, authorization failure).
    \item \textbf{Tool server errors:} The agent cannot successfully execute the tool because the tool server is unavailable, experiencing a failure, or there are other server-side issues. Often these are related to the response code $5XX$
\end{itemize}

We categorize the tool invocation with no errors scenario when the agent successfully calls the correct tool with the appropriate input payload, receives valid tool execution output, and processes the tool output.

\subsubsection{Action Recommendations}
The taxonomy-driven approach not only categorizes the root causes of tool invocation failures but also enables actionable repair strategies. Each error class is linked to a structured set of recommendations, ranging from refinements in parameter descriptions to modifications in tool definitions and backend constraints, by identifying issues related to tool descriptions, parameters, access, or backend servers. Based on the identified errors, Table~\ref{table:error recommendations} in the Appendix provides template-based recommendations for improvement. As shown in the table for the \textit{Missing Parameters} error category, we suggest modifying tool parameter descriptions. When a tool input requires a specific format for a date and the parameter description does not specify it, an incorrect date value is generated, resulting in \textit{Parameter Value Mismatch} error. Additionally, a single input parameter can result in both \textit{Parameter Type Mismatch} and \textit{Parameter Value Mismatch} errors. These recommendations can thus support `Tool Builder' in refining tools and minimizing such failures.

\section{Evaluation}
In this section, we evaluate different components of our framework:
\begin{itemize}
   \item \textbf{NL test case generation}: Evaluate two different models for utterance generation from conventional test cases.
   \item \textbf{Test Execution}: Execute test cases and evaluate error taxonomy of agents with two different models.
   \item \textbf{Data-aware test case generation}: Evaluate the impact of data-aware test case generation on the error taxonomy.
     
\end{itemize}

\begin{table}[t]
\centering
\small
\begin{tabular}{|l|c|c|c|}
\hline
\textbf{Application} & \textbf{\# Tools} & \textbf{\# Input Parameters} & \textbf{\# Test Cases} \\
\hline
DnB         & 22 & 1--13  & 269 \\ \hline
Salesloft   & 36 & 0--41  & 583 \\ \hline
Servicenow  & 39 & 0--14  & 572 \\ \hline
Jira        & 7  & 1--6   & 127 \\ \hline
Kubernetes  & 48 & 0--7   & 418 \\ \hline
Salesforce  & 37 & 0--15  & 442 \\ \hline
\end{tabular}
\caption{Dataset summary}
\label{table:Data set summary}
\end{table}

\subsection{Experimental Setup}

\noindent\textbf{Dataset:}

Our framework is evaluated using a dataset that comprises $189$ enterprise tools. These tools come from from different applications related to business data and analytics (Dun and Bradstreet\footnote{\url{https://developer.dnb.com/}}), ticketing systems (ServiceNow\footnote{\url{https://developer.servicenow.com/dev.do}}),  and sales engagement (Salesloft\footnote{\url{https://developers.salesloft.com/docs/api/}}) etc.,  
Tools vary in complexity in terms of input parameters, response output parameters, and authorization methods. The details are presented in Table~\ref{table:Data set summary}. Many tools have more than 10 input parameters and complex nested output schemas.
In this section, we present results from test case execution. There are a total of $2411$ test cases generated for $189$ tools. For the agentic environment, we used the LangGraph framework to build a React~\cite{yao2023react} agent. 

\noindent{\textbf{Models:}} We use different models for our evaluation: i) API test case generation uses \textit{Mixtral-8x7B-Instruct}~\cite{mixtralexperts}, ii) NL test case generation uses two models 
 - \textit{Mixtral-8x7B-Instruct}~\cite{mixtralexperts} for prompting and  \textit{Granite-3B-Code-Base} model\footnote{\url{https://huggingface.co/ibm-granite/granite-3b-code-base}} for fine-tuning, iii) The  React~\cite{yao2023react} agent uses two different LLMs: \textit{Granite-3.3-8B-Instruct}\footnote{\url{https://huggingface.co/ibm-granite/granite-3.3-8b-instruct}}, \textit{Llama-3.3-70B-Instrcut}\footnote{\url{https://huggingface.co/meta-llama/Llama-3.3-70B-Instruct}} that support tool calling.

\noindent\textbf{Evaluation Metrics:}
We compute the \textbf{error rate per test case} as the fraction of test cases where an error class is observed. A test case can result in multiple error classes. Further, a test case is counted only once, even if the test case has multiple occurrences of the same error. Values lie in the range $[0,1]$, where 1 would indicate that the class of error occurred in every test case. Also, we compute the percentage of test cases that were executed successfully with no errors.

\subsection{Evaluation of NL Test Case Generation}

The process of NL test case generation entails the creation of natural, human-like sentences that would be used to invoke and test a tool in the agentic flow. In the previous section, we discussed the two methods for generating NL test cases. The first is a prompt-based method using an LLM (\textit{Mixtral-8x7B-Instruct}), the prompt used to generate the test cases is presented in Appendix(\ref{appendix:nlgenprompt}). The second, to create more fluent and natural-sounding NL test cases, we used LoRA~\cite{lora2022} to fine-tune \textit{Granite-3B-Code-Base} model. We curated the dataset by combining 400 simple APIs from BFCLv3~\cite{berkeley-function-calling-leaderboard}, with an additional 120 human-curated domain-specific API invocations. The simple APIs from BFCL contain a maximum of 10 parameters each. To better align with real-world enterprise APIs and enhance accuracy, we have designed these domain-specific APIs with more parameters. For the LoRA fine-tuning process, 70\% of the examples from the randomly combined data were used for training, while 15\% each were  allocated to both the validation and test sets.

To evaluate the quality of generated NL test cases, we set up a prompted LLM as a judge with \textit{Llama-3.1-405B-Instruct}\footnote{\url{https://huggingface.co/meta-llama/Llama-3.1-405B-Instruct}} to score them on two criteria: i) Accuracy (Rating: Accurate / Partially Accurate / Inaccurate), and ii) Naturalness (range 1-5), assesses how human-like and natural the NL test case is as opposed to robotic resembling parameter-by-parameter descriptions. The LLM as a judge was validated against human judgment for 15\% of the test set. The human agreement with the rating was found to be 89\% and 93\% for accuracy and naturalness scores, respectively. 

NL test cases that list the parameters are relatively easy for an agent to act upon, but these do not depict the type of NL utterances that users would provide. Hence, naturalness is an important metric to consider for generating NL test cases. (Examples of natural and synthetic NL test cases presented in Appendix:~\ref{appendix:NLExamples})

 \begin{figure}[ht]
    \centering
    \includegraphics[width=0.45\textwidth]{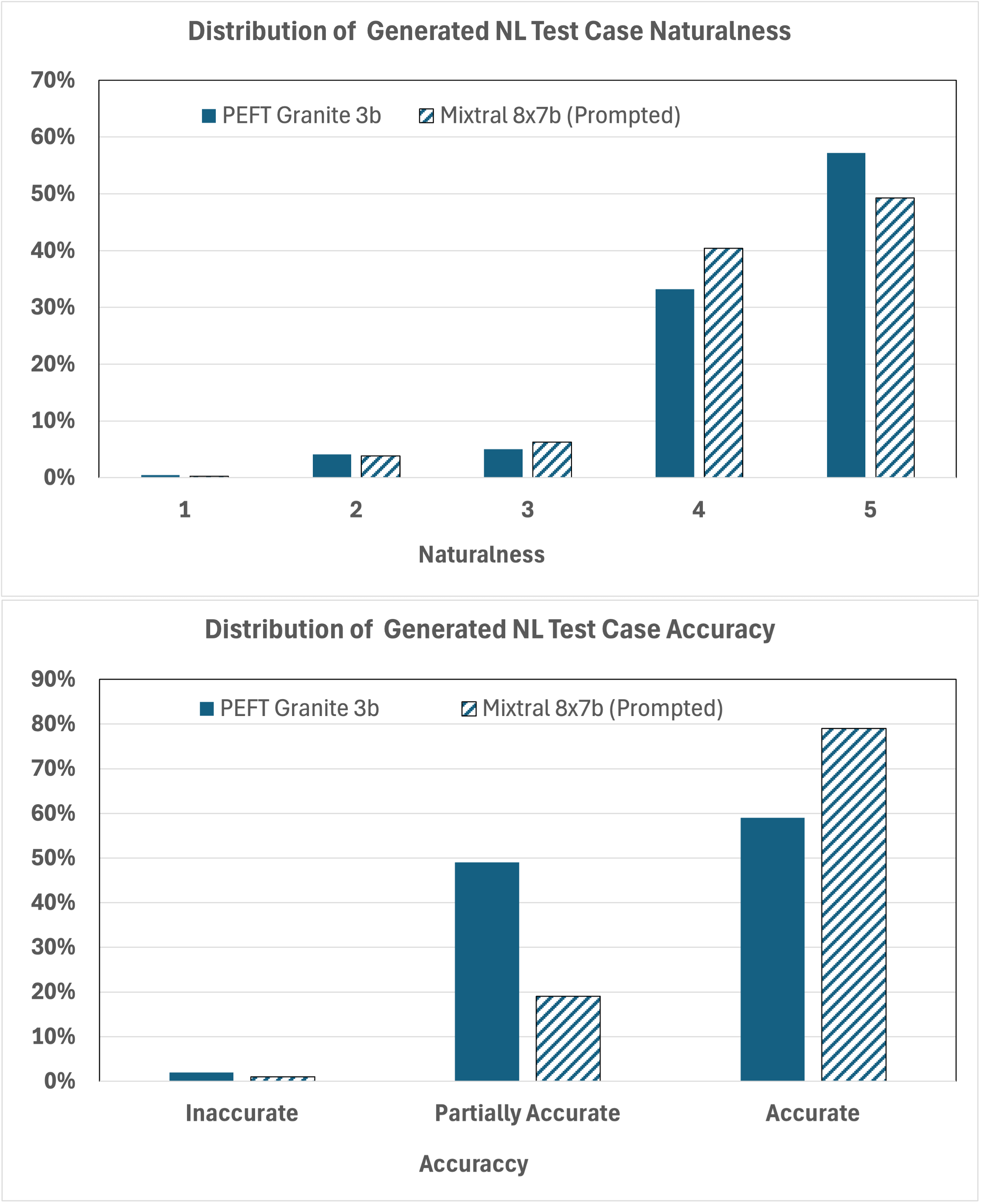} 
    \caption{Distribution of NL test-case accuracy, naturalness on test dataset}
    \label{fig:llm_judge2}
\end{figure}   

 \begin{figure}[ht]
    \centering
    \includegraphics[width=0.45\textwidth]{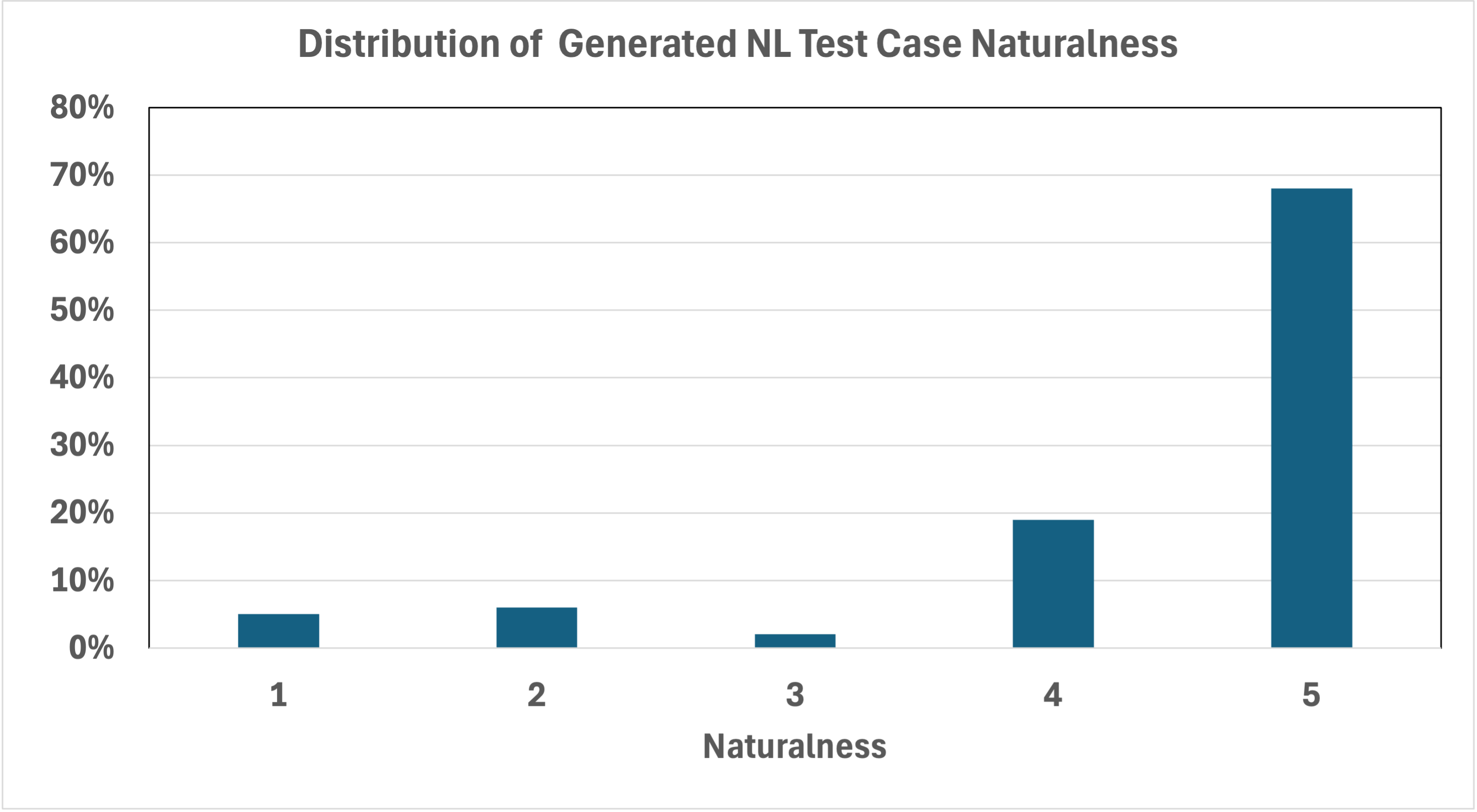} 
    \caption{Distribution of Naturalness for Generated Test Cases}
    \label{fig:llm_judge3}
\end{figure}   

Figures~\ref{fig:llm_judge2} show the distribution of accuracy and naturalness scores of the NL test cases generated by each of the methods. From the graphs, it is evident that the naturalness of NL test cases from PEFT \textit{Granite-3B-Code-Base} model is higher, whereas the accuracy of NL test cases generated from the prompted \textit{Mixtral-8x7B-Instruct} model is higher.

For the evaluation, we use the prompt-based method, though both approaches remain viable depending on the requirements. We generated NL utterances for 2400+ test cases and evaluated them with LLM-as-a-Judge. Overall, 96.4\% were accurate, with only 1.7\% incorrect. Naturalness scores were strongly skewed toward higher ratings, with ~70\% at level 5, ~20\% at level 4, and less than 10\% below 3. Notably, 87.4\% of utterances scored above 3 in naturalness while also being accurate, underscoring both fluency and correctness.

\vspace{1pt}
\noindent\fbox{%
    \parbox{0.45\textwidth}{%
    \textbf{Summary:}  Our analysis shows that natural, human-like utterances can be generated by either a smaller performance-efficient fine-tuned model or a larger model. The NL test generation module selectively retains only test cases that are accurate and with high naturalness.
    }%
}
\subsection{Evaluation of Tool Performance with Agent LLM Models}
For evaluation, we filtered the generated test cases using the `Naturalness', `Accuracy' metrics in Section~\ref{NL_testcase_generation} and retained  `2411' test cases which are `Accurate' and have high naturalness score ($\geq 4$). These NL test cases are executed using two LLMs \textit{Granite-3.3-8B-Instruct} and \textit{Llama-3.3-70B-Instruct} using the ReAct \cite{yao2023react} agent with the LangGraph framework. Table~\ref{table:error-analysis} shows different error categories of the error taxonomy and their percentage per test case. Approximately 30 percent of test cases execute without any errors. 

The majority of the tool execution errors are due to incorrect tool parameter type and value. In the below example, the expected tool input payload should contain the parameter \texttt{per\_page} with value 100 as an integer type, and \texttt{sort} with value `DESC' as a string type. However, the tool is invoked with the incorrect type for the parameter \texttt{per\_page}, and an incorrect value for the parameter \texttt{sort}. Both the LLMs suffered from recursive tool invocation and output mismatch errors due to insufficient tool definition. 
\vspace{2pt}
\noindent\fbox{%
    \begin{small}
    \parbox{0.45\textwidth}{%
    \textbf{Example of incorrect parameter type and value} \\
    \textit{Expected tool input payload}: \\
    \texttt{\{`per\_page':100,
    `sort':`DESC'\}}\\
    \textit{In-correct tool input payload}: \\
    \texttt{\{`per\_page':`100',
    `sort':`DESC order'\}}

    }%
    \end{small}
}

For a few test cases, an incorrect tool selection error is observed, and some test cases fail because of incorrect tool access or tool service unavailability. Several test cases produced an Empty output error due to invalid input values provided in the test case. We observed across LLM models that the performance of the tool calling varies, but the distribution of errors is similar. Tool errors related to parameter type and value are more prevalent, and to fix these types of errors, we recommend actions shown in Table~\ref{table:error recommendations} in the Appendix. 

\vspace{2pt}
\noindent\fbox{%
    \parbox{0.45\textwidth}{%
    \textbf{Summary:} LLMs exhibit significant and varying errors when interacting with enterprise tools. A major source of these errors, observed in 13-19\% of test cases, stems from issues in generating correct input payloads, specifically with respect to parameter types, values.
    }%
}

\begin{table}[]
\small
\centering
\begin{tabular}{|cc|c|c|}
\hline
\multicolumn{1}{|c|}{\textbf{Category}}                                                               & \textbf{Error Type}                                                 & \textbf{\begin{tabular}[c]{@{}c@{}}Llama-\\ 3.3-70B-\\ Instruct\end{tabular}} & \textbf{\begin{tabular}[c]{@{}c@{}}Granite-\\ 3.3-8B-\\ Instruct\end{tabular}} \\ \hline
\multicolumn{1}{|c|}{\multirow{3}{*}{\begin{tabular}[c]{@{}c@{}}Tool \\ Invocation\end{tabular}}}     & \begin{tabular}[c]{@{}c@{}}Tool Not \\ Identified\end{tabular}      & 0.02                                                                          & 0.01                                                                           \\ \cline{2-4} 
\multicolumn{1}{|c|}{}                                                                                & \begin{tabular}[c]{@{}c@{}}Incorrect Tool \\ Selection\end{tabular} & 0.02                                                                          & 0.08                                                                           \\ \cline{2-4} 
\multicolumn{1}{|c|}{}                                                                                & \begin{tabular}[c]{@{}c@{}}Repeated Tool \\ Invocation\end{tabular} & 0.11                                                                          & 0.16                                                                           \\ \hline
\multicolumn{1}{|c|}{\multirow{4}{*}{\begin{tabular}[c]{@{}c@{}}Input \\ Specification\end{tabular}}} & \begin{tabular}[c]{@{}c@{}}Parameter Value \\ Mismatch\end{tabular} & 0.17                                                                          & 0.13                                                                           \\ \cline{2-4} 
\multicolumn{1}{|c|}{}                                                                                & \begin{tabular}[c]{@{}c@{}}Parameter Type \\ Mismatch\end{tabular}  & 0.19                                                                          & 0.03                                                                           \\ \cline{2-4} 
\multicolumn{1}{|c|}{}                                                                                & \begin{tabular}[c]{@{}c@{}}Incorrect \\ Parameter\end{tabular}      & 0.16                                                                          & 0.15                                                                           \\ \cline{2-4} 
\multicolumn{1}{|c|}{}                                                                                & \begin{tabular}[c]{@{}c@{}}Missing \\ Parameter\end{tabular}        & 0.10                                                                          & 0.15                                                                           \\ \hline
\multicolumn{1}{|c|}{\multirow{4}{*}{\begin{tabular}[c]{@{}c@{}}Output \\ Processing\end{tabular}}}   & \begin{tabular}[c]{@{}c@{}}Exceeding \\ Token Limit\end{tabular}    & 0.00                                                                          & 0.00                                                                           \\ \cline{2-4} 
\multicolumn{1}{|c|}{}                                                                                & Malformed Output                                                    & 0.00                                                                          & 0.02                                                                           \\ \cline{2-4} 
\multicolumn{1}{|c|}{}                                                                                & Output Mismatch                                                     & 0.08                                                                          & 0.08                                                                           \\ \cline{2-4} 
\multicolumn{1}{|c|}{}                                                                                & Empty Output                                                        & 0.14                                                                          & 0.11                                                                           \\ \hline
\multicolumn{1}{|c|}{\multirow{2}{*}{\begin{tabular}[c]{@{}c@{}}Tool \\ Execution\end{tabular}}}      & Tool Access Error                                                   & 0.04                                                                          & 0.04                                                                           \\ \cline{2-4} 
\multicolumn{1}{|c|}{}                                                                                & Tool Server Error                                                   & 0.01                                                                          & 0.01                                                                           \\ \hline
\multicolumn{2}{|c|}{Fraction with no Errors}                                                                                                                               & 0.26                                                                          & 0.30                                                                           \\ \hline
\end{tabular}
\caption{Error rate per test case observed across LLMs.}
\label{table:error-analysis}
\end{table}

\vspace{-5pt}
\subsection{Evaluation of Data-Aware Testing}
For a subset of $705$ test cases for which the tool graph is approved by the `Tool Builder'. We compare the test cases results with and without the data-aware inputs. The objective of data-aware testing is to provide relevant test inputs such that tool output related errors are minimised.

In Table \ref{tab:data_aware_test_case}, we observe that the error rate of \emph{empty tool outputs} as well as instances of \emph{output mismatches} is reduced substantially when using the data-aware inputs compared to the without data-aware baseline. We observed that data-aware testing also improves the fraction of test cases with no errors.This finding demonstrates two important outcomes. First, data-aware testing improves the \emph{validity} of the test cases by aligning inputs with the syntactic and semantic requirements of each tool. Second, it enhances the \emph{utility} of those cases by ensuring that tool executions yield non-trivial and meaningful outputs. Hence, these improvements show that incorporating knowledge of tool definition docstrings and inter-tool dependencies enables the generation of test cases that are relevant.

\begin{table}[]
\centering
\small
\begin{tabular}{|cc|cc|}
\hline
\multicolumn{1}{|c|}{\multirow{2}{*}{\textbf{Category}}}                                           & \multirow{2}{*}{\textbf{Error Type}}                             & \multicolumn{2}{c|}{\textbf{\begin{tabular}[c]{@{}c@{}}Per test case\\ (percentage)\end{tabular}}}                                             \\ \cline{3-4} 
\multicolumn{1}{|c|}{}                                                                             &                                                                  & \multicolumn{1}{c|}{\begin{tabular}[c]{@{}c@{}}Without\\ data-aware\end{tabular}} & \begin{tabular}[c]{@{}c@{}}With data-\\ aware\end{tabular} \\ \hline
\multicolumn{1}{|c|}{\multirow{4}{*}{\begin{tabular}[c]{@{}c@{}}Output\\ Processing\end{tabular}}} & \begin{tabular}[c]{@{}c@{}}Exceeding \\ Token Limit\end{tabular} & \multicolumn{1}{c|}{0.01}                                                         & 0.00                                                       \\ \cline{2-4} 
\multicolumn{1}{|c|}{}                                                                             & Malformed Output                                                 & \multicolumn{1}{c|}{0.00}                                                         & 0.02                                                       \\ \cline{2-4} 
\multicolumn{1}{|c|}{}                                                                             & Output Mismatch                                                  & \multicolumn{1}{c|}{0.43}                                                         & \textbf{0.34}                                              \\ \cline{2-4} 
\multicolumn{1}{|c|}{}                                                                             & Empty Output                                                     & \multicolumn{1}{c|}{0.26}                                                         & \textbf{0.16}                                              \\ \hline
\multicolumn{2}{|c|}{Fraction with no Errors}                                                                                                                         & \multicolumn{1}{c|}{0.18}                                                         & \textbf{0.24}                                              \\ \hline
\end{tabular}
\caption{Comparison of with and without data-aware testing}
\label{tab:data_aware_test_case}
\end{table}
\vspace{-3pt}

\vspace{5pt}
\noindent\fbox{%
    \parbox{0.45\textwidth}{%
    \textbf{Summary:} Data-aware testing reduces empty output, and output mismatch errors by 10\%  producing relevant, valid test cases for multi-tool workflows. 
    }%
}

\section{Deployment Details}

Our framework is deployed as part of the alpha release of the IBM Watsonx Orchestrate Agent Development Kit (ADK)\footnote{\url{https://developer.watson-orchestrate.ibm.com/}}, specifically designed for the tool builder persona. The Agent Development Kit (ADK) is a set of Command Line Interface (CLI) utilities and Python-based library with a tooling environment where builders can configure, deploy, and manage agents and tools within Watsonx Orchestrate. As part of the ADK, our framework enables a tool builder to test a tool with an agent. Currently, the support is provided through a CLI where a user can \texttt{generate} test cases for a tool and further \texttt{evaluate} the tool with an agent. The error taxonomy, along with the relevant recommendations, is provided to the user. To date, builders have wrapped more than 600 APIs as tools that span a wide range of domains, including IT, human resources, finance, procurement, and sales.

Before our framework was introduced, builders manually created tools and tested only 5–6 test cases for each tool. This process was slow, labor-intensive, and left significant coverage gaps. Our automated test case generation feature eliminates this bottleneck by producing comprehensive test cases at scale, improving coverage while reducing manual effort in creating and testing the tools. 
Beyond efficiency gains, our framework effectively helps identify tools that require attention from builders. Analysis of test-case pass rates (i.e., fraction of test cases that pass for each tool) shows that 34\% of tools fall in the 0–25\% pass rate range, indicating the need for substantial iteration. Another 30\% of tools, with pass rates between 25–75\%, require moderate iteration, while 36\% of tools achieve 75–100\% pass rates and are already in good working condition. These insights help prioritize builder effort on relevant tools.

Our analysis of Tool Builder effort shows that this automation, along with actionable recommendations, has reduced build effort per tool by 30\%—equivalent to approximately two days saved per tool—resulting in over 1,200 person-days saved to date, with the number continuing to grow.
Additionally, the generated test cases are used for continuous and regression testing of tools to ensure proper functioning of updated tools. For releasing new versions of the testing framework with new features and enhancements, we have integrated a DevOps pipeline to support release management.

\section{Conclusion and Future Work}
In this paper, we present a framework that adapts existing API testing methods to evaluate `agent-ready' enterprise tools. The framework utilizes an LLM to generate natural language test cases for tools and leverages tool testing execution reports to identify and categorize failures within an error taxonomy, offering recommendations for improvement. In the future, we plan to extend our work to generate multi-turn utterances that simulate a conversation leading to tool invocation and handling multi-step interactions involving sequential tool calls.Additionally, the error taxonomy related to processing sequential tool calls will be explored. 

\bibliography{aaai2026}
\newpage
\appendix
\section{Appendix}\label{Appendix}

\subsection{Test Case Generation}\setlabel{Test Case Generation}{appendix:testcasegenprompt} 
Following is the sample prompt that is used to generate the test case with all parameters:

\noindent\fbox{%
    \parbox{0.47\textwidth}{%
    \texttt{Given a function definition with a set of parameters, your task is to create a single, comprehensive test case. You must provide realistic values for all parameters, without using dummy words or None. If example values are provided for any parameter, use only those values. The test case should be presented in a dictionary format, with each parameter and its corresponding value included. Ensure that all parameters are accounted for and no parameters are excluded. 
    Function definition: \{`tool name':*** , `tool inputs': ***, `tool outputs': ***\}\\
    Output:
    }}%
}

Following is the sample prompt that we used to generate the test case with LLM selected subset of parameters:

\noindent\fbox{%
    \parbox{0.47\textwidth}{%
    \texttt{As a test case generation expert, create a variety of test cases for the given function definition. Each test case should include all mandatory parameters and a mix of valid optional parameters. If example values are provided for a parameter, restrict your test cases to those values. For parameters without specified values, generate realistic ones. Avoid using placeholder words like 'example' or None as parameter values. Present the test cases in dictionary format.\\
    Function definition: \{`tool name':*** , `tool inputs': ***, `tool outputs': ***\}\\
    Output:
    }}%
}

\subsection{NL Utterance Generation} 
\setlabel{NL Utterance Generation}{appendix:nlgenprompt}Following is the sample prompt that we used to generate the natural language:

\noindent\fbox{%
    \parbox{0.47\textwidth}{%
    \texttt{Given the following function and input parameters, what should 
be the natural language utterance?\\
Function definition: \{`tool name':*** , `tool inputs': ***, `tool outputs': ***\}\\
Input parameters: {payload}\\
Output:
    }}%
}

\subsection{NL Utterance Evaluation}\setlabel{NL Utterance Evaluation}{appendix:NL-LaJ} The prompt presented below is employed to evaluate the degree of naturalness and the correctness of the generated NL utterance using LLM as a Judge.

\noindent\fbox{%
    \parbox{0.47\textwidth}{%
\texttt{You are provided with a Function definition, a sample  payload, and an utterance. Your task is to evaluate the utterance and 
provide the following two ratings}. \\

\texttt{1. Naturalness: Rating from 1 to 5.  describes how human-like the sentence is as per the examples below. Natural and 
human-like sentences get a higher rating.} \\
\texttt{In-context examples for naturalness rating}\\
\texttt{2. Accuracy (Rating: Accurate / Partially Accurate / Inaccurate)\\
Determine how well the utterance reflects the information in the sample payload:}\\

\texttt{In-context examples for accuracy rating}\\
\texttt{Provide the ratings for
Function definition: \{`tool name':*** , `tool inputs': ***, `tool outputs': ***\}\\
Sample Payload: {payload}\\
utterance: {utterance}
    }}%
}

\subsection{Examples of Natural and Synthetic NL Utterances}\setlabel{Natural and Synthetic NL Utterances}{appendix:NLExamples}
The following are examples of natural and synthetic NL utterances

\noindent\fbox{%
    \parbox{0.47\textwidth}{%
     \begin{small}
    \textbf{Example of a natural sentence: }
Update my contact information. Change my email to jane.doe@fastmail.com, my phone number to +1234567890, and my home phone number to +1234567890. Also, change my work email to alex.mccaw@fastmail.com. My home address is 123 Main St, Anytown, USA 12345. My work address is 456 Elm St, Oakland, CA 94607. For work, my job title is Senior Software Engineer.
\end{small}
    }%
}

\noindent\fbox{%
    \parbox{0.47\textwidth}{%
     \begin{small}
\textbf{Example of a synthetic sentence that is not fluent:} 
Update my requester with time format as 14:14:14, time zone as UTC+04:00, mobile phone number as 1234567890, language as French, primary email as ehuntington@workday.net, work phone number as 2345678901, last name as Rodriguez, ID as 12345, reporting manager ID as JOB-POSTING-6-1227, location ID as Paris, address as 900 Building 2, and first name as Sarah.
\end{small}
    }%
}

\newpage
\subsection{Tool Dependency Graph }
\setlabel{Tool Dependency Graph}{ref:tool_graph}
Following is the sample prompt that we used to identify tool dependency:\\
\noindent\fbox{%
    \parbox{0.47\textwidth}{%
\texttt{You are a software engineering expert. You are given Python tool definitions with docstrings describing their parameters and functionality}.\\
\texttt{Tool definitions: [ tool1, tool2 ]}\\
\texttt{Your task is to find the dependencies for {tool}.
If {tool} needs the output of another function, return their names in a list (e.g., [ToolB, ToolC]).
If not, return None.}\\
\texttt{Instructions:
Check the input/output structure of each function. Review each parameter of {tool} to see if it can be derived from the output of other functions.
Return only the list of required function names or None.
Final Answer is:                   
    }}%
}
\vfill\null
\columnbreak
\subsection{Tool Dependency Graph -  Parameter Mapping}\setlabel{Parameter Mapping}{ref:argument_mapping}

The following is a sample prompt to identify the parameter mapping of the tools in the tool dependency graph:\\
\noindent\fbox{%
  \parbox{0.47\textwidth}{%
    \texttt{Given a docstring for a specific function identified by \{tool\}, your task is to:}

\begin{itemize}
  \item \texttt{Parse each line of the docstring, especially those containing \texttt{:param}, \texttt{:type}, \texttt{:return}, or similar annotations.}
  \item \texttt{Identify and extract the following:}
    \begin{itemize}
      \item \texttt{source\_tool name}
      \item \texttt{source\_param names}
      \item \texttt{target\_param names}
    \end{itemize}
  \item \texttt{Return the results as a JSON-formatted list with clearly labeled fields.}
\end{itemize}

\texttt{You may use the following reference specification block for context:}

\texttt{\textbf{Training Example:}}

\parbox{0.9\linewidth}{\texttt{input: :param article\_id: The id returned by the `fetch\_articles` tool.}}

\parbox{0.9\linewidth}{\texttt{output: [{"source\_tool": "fetch\_articles.py", "source\_param": "article\_id", "target\_param": "article\_id"}]}}

\texttt{Please don't generate irrelevant content. Only provide the answer. Carefully parse each \texttt{:param} line. Keep in mind that Python keywords do not contain spaces. Also, avoid self-loops.}

\texttt{\textbf{Test Example:}}

\texttt{input : \{docstring\}}

\texttt{output:}

  }%
}

\begin{table*}[ht]
\centering
\small
\begin{tabular}{|p{3.8cm}|p{3.8cm}|p{9.2cm}|}
\hline
\textbf{Error Category} & \textbf{Area of Concern} & \textbf{Recommended Guidance} \\ \hline

\multicolumn{3}{|c|}{\textbf{Tool Definition \& Documentation}} \\ \hline
Tool Invocation – Tool Not Identified &
Definition Clarity &
The agent could not match any tool, though the expected tool was \{TOOL\_NAME\}.  
Consider refining the description of \{TOOL\_NAME\} to make its purpose clearer and more discoverable. \\ \hline

Tool Invocation – Incorrect Tool Selection &
Definition Clarity &
The agent selected the wrong tool. Expected: \{TOOL\_NAME\}, but invoked: \{TOOL\_NAME\}.  
You may improve the description of the correct tool or highlight distinctions between overlapping tools. \\ \hline

Input Specification – Missing Parameters &
Input Schema &
Some required inputs (e.g., \{PARAMETER\_1, PARAMETER\_2\}) were not supplied to \{TOOL\_NAME\}.  
Ensure parameter descriptions are prominent and include defaults/examples if possible. \\ \hline

Input Specification – Incorrect Parameters &
Input Schema &
The agent hallucinated unsupported inputs (e.g., \{PARAMETER\_1, PARAMETER\_2\}).  
Clarify the input schema for \{TOOL\_NAME\}. \\ \hline

Input Specification – Parameter Type Mismatch &
Input Schema &
\{TOOL\_NAME\} received parameters with invalid types.  
Explicitly document accepted types for \{PARAMETER\_1, PARAMETER\_2\}, or add preprocessing checks. \\ \hline

Input Specification – Parameter Value Mismatch &
Input Schema &
\{TOOL\_NAME\} received values outside the expected range.  
Provide examples of valid values and constraints. \\ \hline

Output Processing – Malformed Output &
Output Schema &
The agent could not parse \{TOOL\_OUTPUT\} from \{TOOL\_NAME\}.  
Ensure the output schema is consistent and provide structured examples. \\ \hline

Output Processing – Output Mismatch &
Output Schema &
The output from \{TOOL\_NAME\} lacked required fields (e.g., \{PARAMETER\_1, PARAMETER\_2\}).  
Refine the tool’s response format to align with the tool definition. \\ \hline

\multicolumn{3}{|c|}{\textbf{Tool Implementation}} \\ \hline
Tool Invocation – Recursive Tool Invocation &
Implementation Logic &
\{TOOL\_NAME\} was invoked repeatedly because the model could not interpret its output \{TOOL\_OUTPUT\}.  
Consider refining the output format or post-processing logic to break recursion. \\ \hline

Output Processing – Empty Output &
Implementation Coverage &
The tool \{TOOL\_NAME\} returned no results for the given inputs.  
Verify with data-aware test cases. \\ \hline

\multicolumn{3}{|c|}{\textbf{Agent–Tool Interaction}} \\ \hline
Output Processing – Exceeding Token Limit &
LLM Constraints &
The output of \{TOOL\_NAME\} exceeded the token limit of \{LLM\_MODEL\}.  
Consider truncation, summarization, or using a model with a higher context window. \\ \hline

\multicolumn{3}{|c|}{\textbf{Execution \& Configurations}} \\ \hline
Tool Execution – Access Error &
Authentication &
Access credentials for \{TOOL\_NAME\} are invalid.  
Recheck authentication settings and ensure keys/tokens are active. \\ \hline

Tool Execution – Server Error &
Service Reliability &
\{TOOL\_NAME\} is temporarily unavailable due to a server error.  
Retry after some time, or implement fallback handling in case of downtime. \\ \hline

\end{tabular}
\caption{Categorized error taxonomy and action recommendation across tool definition, implementation, interaction, and execution.}
\label{table:error recommendations}
\end{table*}


\begin{figure*}
\centering

\begin{lstlisting}[language=Python]
import json
import requests
from typing import *
from langchain_core.tools import tool

@tool
def deleteCoreV1NamespacedPod(namespace: str, name: str, dryRun: Optional[str] = None, gracePeriodSeconds: Optional[int] = None, orphanDependents: Optional[bool] = None, propagationPolicy: Optional[str] = None, requestBody: Optional[dict] = None):
	""" Deletes a specified Pod within a given namespace. This operation allows for various options such as performing a dry run to simulate the deletion, setting a grace period before deletion, and specifying how dependent objects should be handled. The operation supports parameters to control the deletion process, including whether to orphan dependent objects or manage their deletion in the background or foreground.

	:param namespace: The namespace in which the pod resides. This parameter is required to identify the specific namespace where the pod to be deleted is located.
	:param name: The name of the pod to be deleted.
	:param dryRun: Indicates whether the operation should be performed as a dry run. If set to 'All', all dry run stages will be processed. An invalid or unrecognized value will result in an error response and no further processing of the request.
	:param gracePeriodSeconds: The duration in seconds before the pod is deleted. A value of zero indicates immediate deletion. If not specified, the default grace period for the pod type will be used.
	:param orphanDependents: Indicates whether dependent objects should be orphaned. If set to true, the 'orphan' finalizer will be added to the object's finalizers list. If set to false, it will be removed. Note: This field is deprecated; use PropagationPolicy instead. This field will be removed in version 1.7.
	:param propagationPolicy: Specifies the garbage collection policy for the pod's dependents. Acceptable values are: 'Orphan' - orphan the dependents; 'Background' - delete the dependents in the background; 'Foreground' - delete all dependents in the foreground. Either this field or OrphanDependents may be set, but not both. The default policy is determined by the existing finalizer set in the metadata.finalizers and the resource-specific default policy.
	:return: The JSON response from the API. 

	Input Example:
	namespace = 'default'
	name = 'web-server-1'
	dryRun = 'All'
	orphanDependents = True
	propagationPolicy = 'Orphan'
	requestBody = {'apiVersion': 'v1', 'dryRun': ['All'], 'gracePeriodSeconds': 0, 'kind': 'Pod', 'orphanDependents': True, 'propagationPolicy': 'Orphan'}
	"""

	header = {
		'accept': 'application/json',
		'content-type': 'application/x-www-form-urlencoded'
	}
	queryParam = {'dryRun' : dryRun, 'gracePeriodSeconds' : gracePeriodSeconds, 'orphanDependents' : orphanDependents, 'propagationPolicy' : propagationPolicy}

	api_url = f"http://localhost:8080/api/v1/namespaces/{namespace}/pods/{name}"
	response = requests.delete(api_url, headers=header, params=queryParam, json=requestBody)
	return {"status_code": response.status_code, "response": response.json()}

\end{lstlisting}
\caption{Python tool by Tool Creation component}\label{fig:kubernetes-tool}
\end{figure*}

\end{document}